\documentclass[twocolumn]{revtex4}  %% REVTeX 4.0
\usepackage{graphicx}
\usepackage{subfigure}
\begin{document}
\title{Integrated magneto-optical traps on a chip}
\author{S. Pollock}
%\email{s.pollock06@imperial.ac.uk}
\author{J. P. Cotter}
\email{j.cotter@imperial.ac.uk}
\author{A. Laliotis}
\author{E. A. Hinds}
\email{ed.hinds@imperial.ac.uk}
\affiliation{The Centre for Cold Matter, Blackett Laboratory, Imperial College London, SW7 2AZ}

\newcommand{\ket}[1]{\vert #1 \rangle}

\begin{abstract}
We have integrated magneto-optical traps (MOTs) into an atom chip by etching pyramids into a silicon wafer.  These have been used to trap atoms on the chip, directly from a room temperature vapor of rubidium. This new atom trapping method provides a simple way to integrate several atom sources on the same chip. It represents a substantial advance in atom chip technology and offers new possibilities for atom chip applications such as integrated single atom or photon sources and molecules on a chip.
\end{abstract}
\maketitle

Atom chips are microfabricated devices that control
electric, magnetic and optical fields in order to trap and
manipulate cold atom clouds \cite{Hinds99}, \cite{Folman02},
\cite{Fortagh07}, and to form Bose-Einstein
condensates \cite{Hansel01}, \cite{Ott01},
\cite{Sinclair05}. Prospective applications include atomic clocks, atom interferometers,
and quantum information processors. Despite enormous progress in the development of atom chips over recent years, the method of loading them has continued to be quite inconvenient. Atoms are first collected in a macroscopic magneto-optical trap (MOT) far from the surface, which requires several laser beams and often uses macroscopic magnetic field coils external to the chip. The atoms are then loaded into a magnetic trap, which finally has to be moved very accurately to transport them to the small structures on the chip. A simple alternative has been proposed \cite{trupke06}, in which the silicon wafer of the chip is processed to make pyramidal hollows wherever atoms are required. A single incident light beam is then multiply reflected in each pyramid to form the set of appropriately-polarized light beams needed for magneto-optical trapping \cite{lee96}. This has the virtue that atoms are cooled and collected directly on the chip at the locations of interest.  References \cite{trupke06,lewis09} report some technical aspects of fabricating integrated MOTs on a chip, but such a source has never previously been demonstrated.

In this letter we present the successful operation of an integrated atom source using a pyramid MOT, lithographically fabricated in silicon, which collects atoms from a background vapor and cools and traps them on the chip. We outline the micro-fabrication process required to produce deep pyramidal structures in silicon, and the steps needed to achieve adequate optical quality of the pyramid facets.  We present data for three different MOTs to establish how the atom number scales with the size of the MOT and we discuss possible future applications.

\begin{figure}[t]
\centering
\subfigure[\label{fig:image1_pic}]{
\includegraphics[width = 0.3\columnwidth]{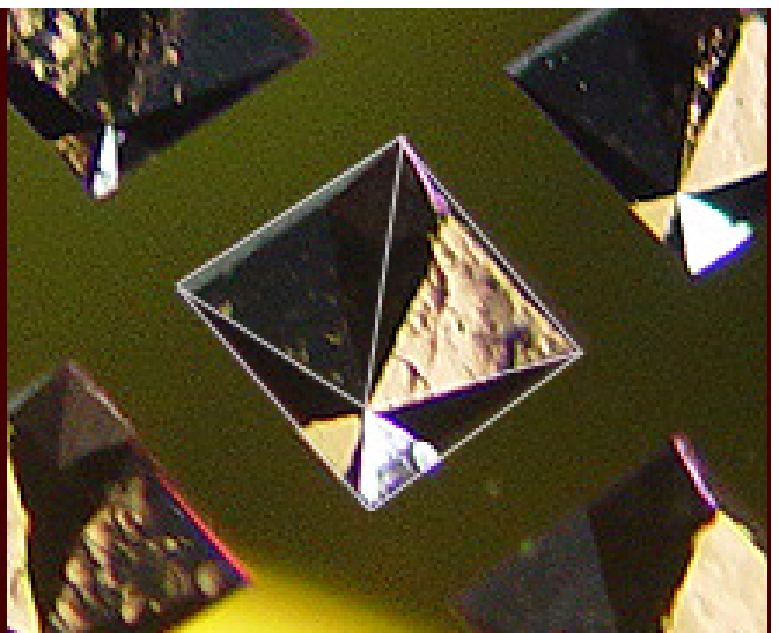}
}
\subfigure[\label{fig:image2_pic}]{
\includegraphics[width = 0.3\columnwidth]{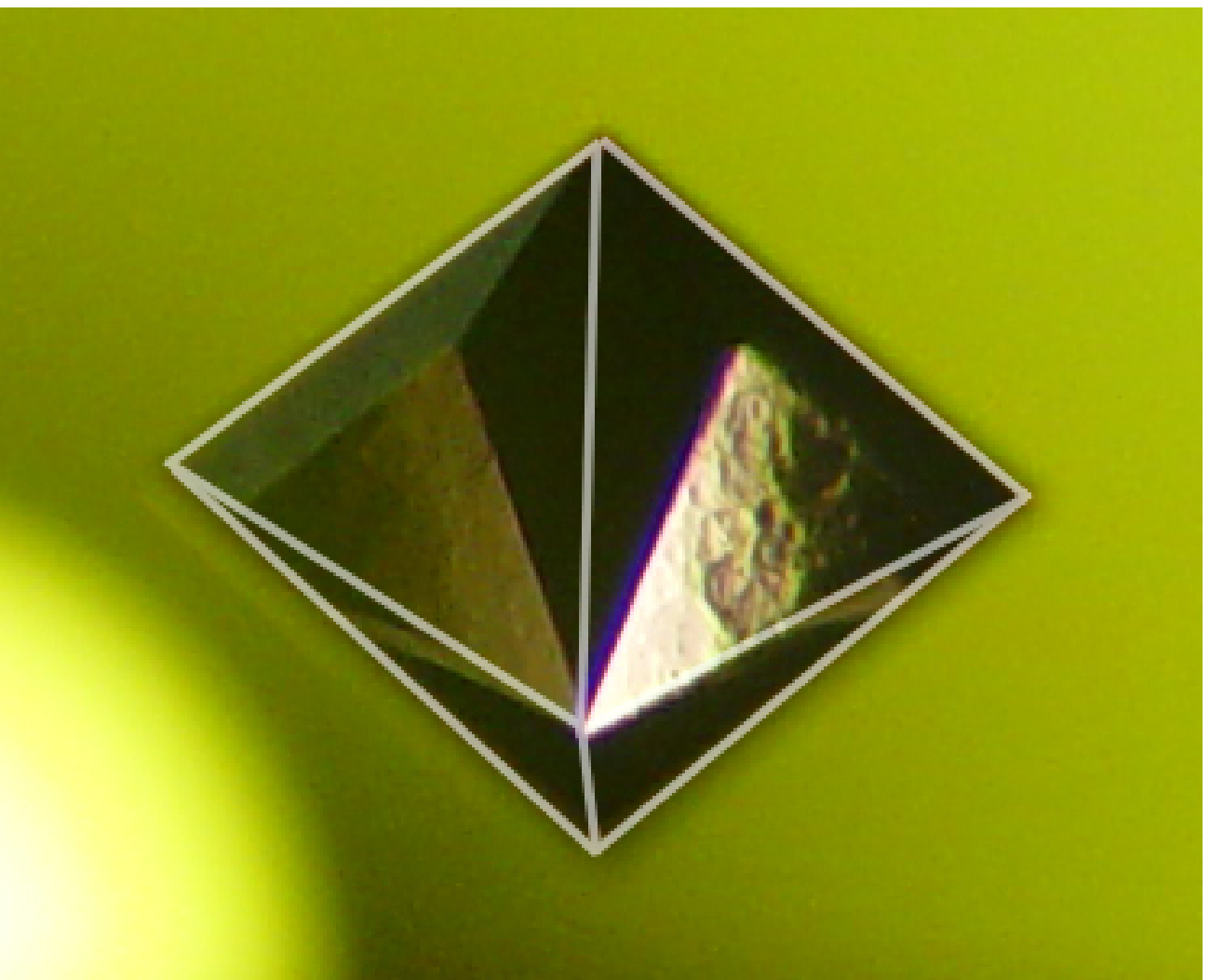}
}
\subfigure[\label{fig:image3_pic}]{
\includegraphics[width = 0.3\columnwidth]{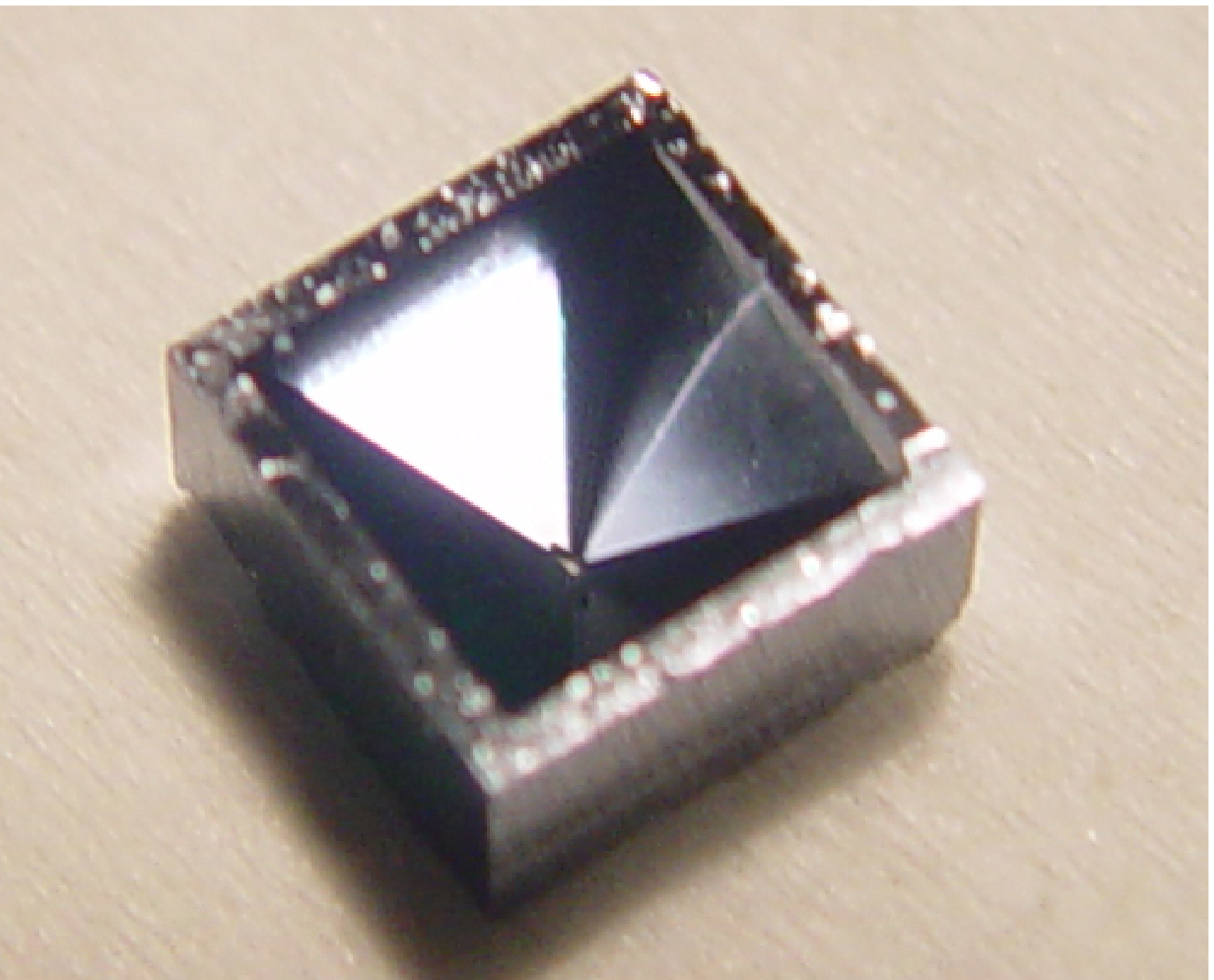}
}
\subfigure[\label{fig:image1_mic}]{
\includegraphics[width = 0.3\columnwidth]{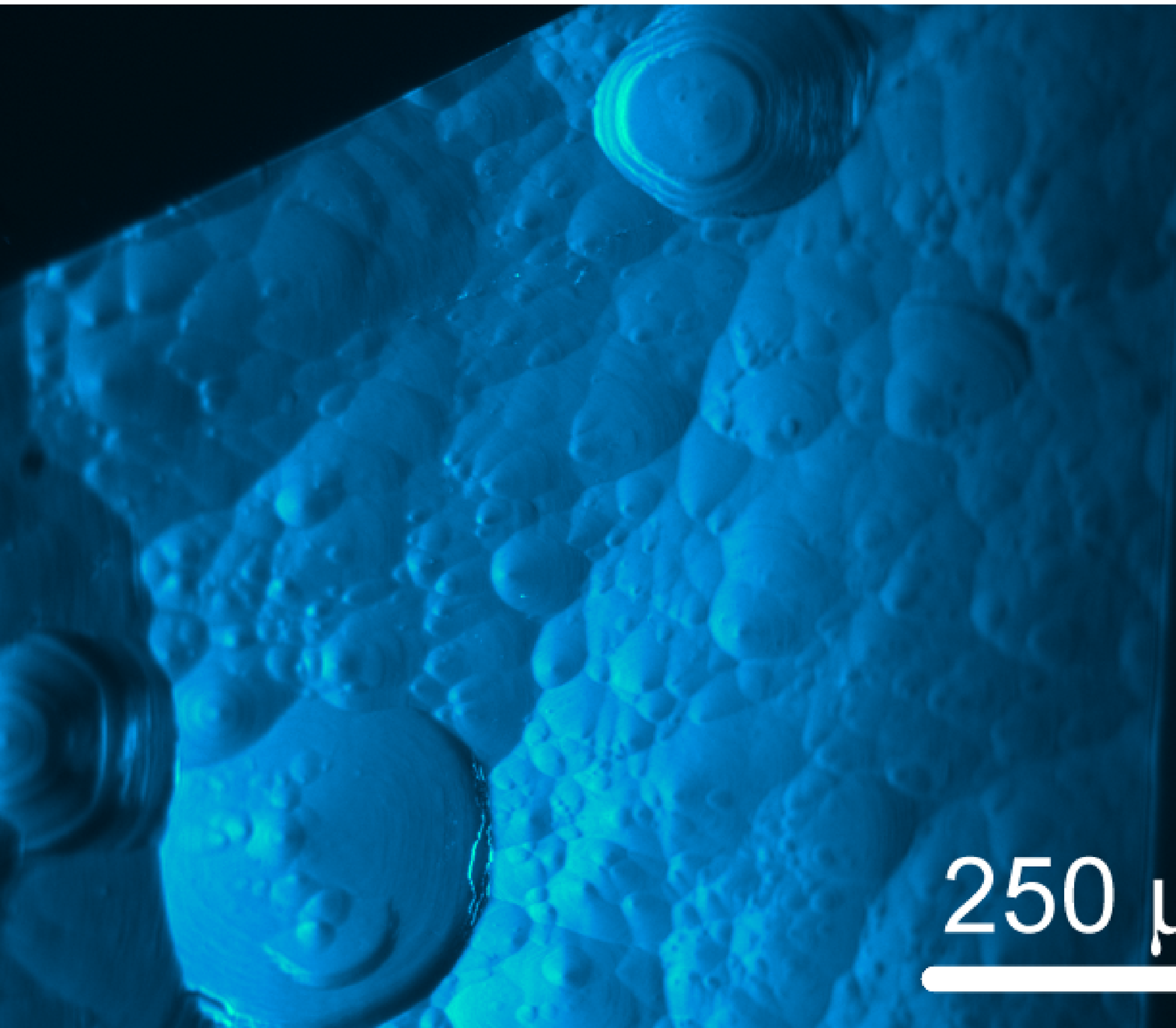}
}
\subfigure[\label{fig:image2_mic}]{
\includegraphics[width = 0.3\columnwidth]{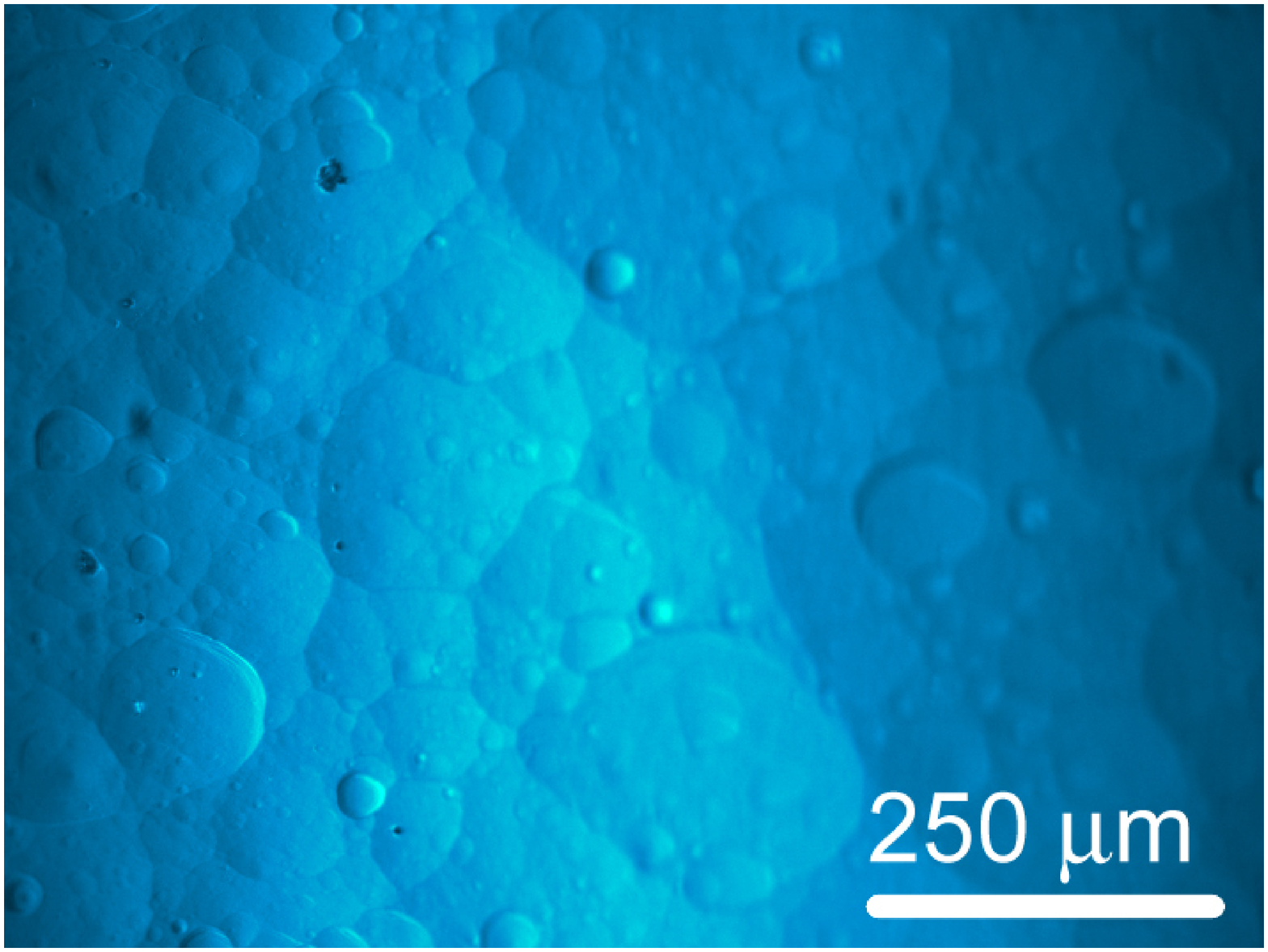}
}
\subfigure[\label{fig:image3_mic}]{
\includegraphics[width = 0.3\columnwidth]{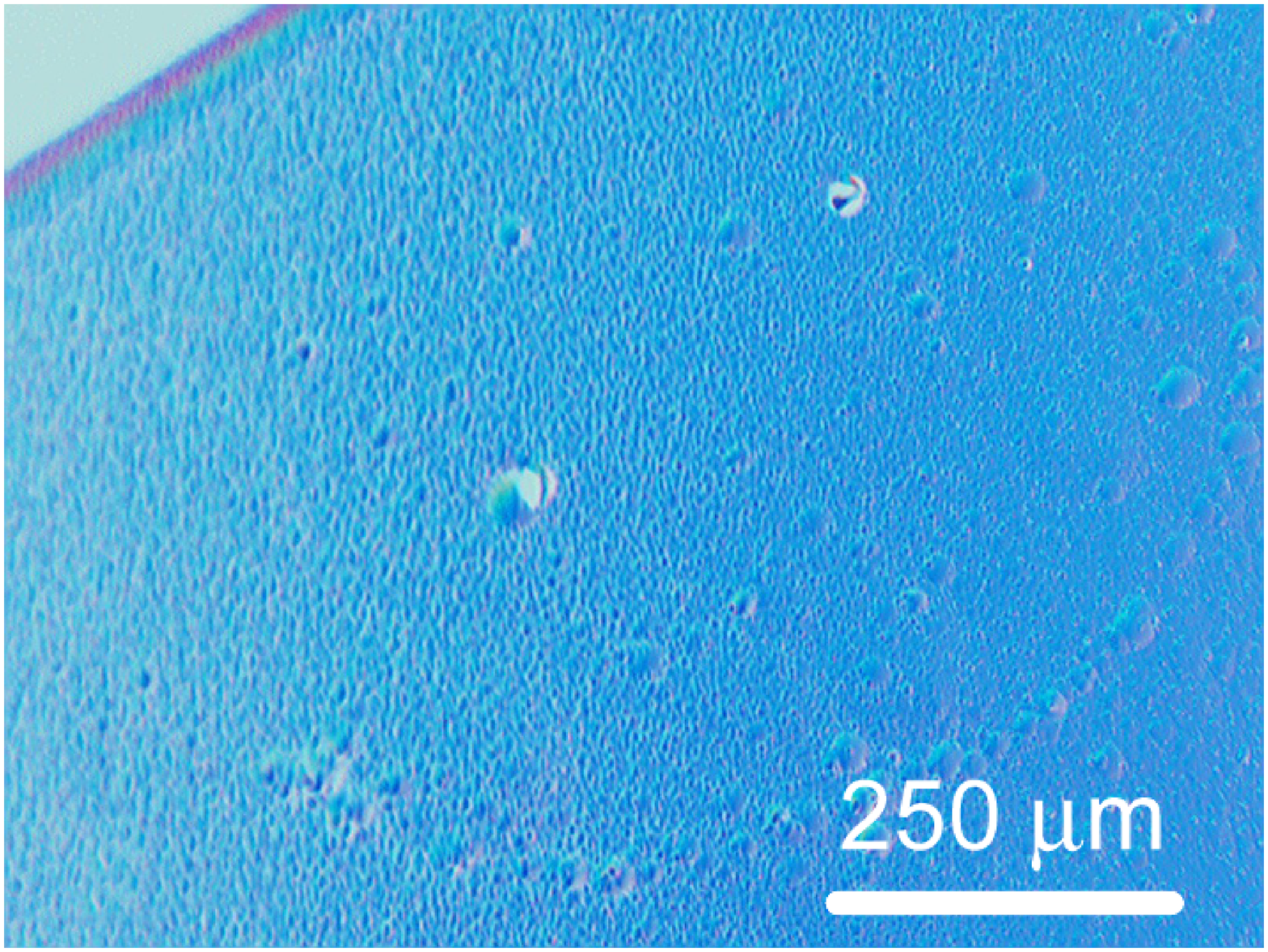}
}
\caption{Pyramids after various stages of processing. 
Top row: photographs. Bottom row: optical microscope images of an internal pyramid face. (a, d) Array of pyramids $2.8\,$mm deep, etched without agitation. (b, e) Single pyramid $2.1\,$mm deep, formed with air bubbling through etchant.  (c, f) Pyramid $2.8\,$mm deep, etched with agitation, then polished by 30 minutes of ICP etching. This particular wafer was diced to make the individual pyramid shown.}
\end{figure}

In order to fabricate deep pyramids, we start with silicon wafers of $3\,$mm thickness, cut on the [100] plane. These are coated on both sides with a thick ($350\,$nm) protective layer of low stress silicon nitride, applied by low pressure chemical vapor deposition.  We use photo-lithography with reactive ion etching to open square apertures in one of the layers, thereby exposing the silicon. Etching in a solution of KOH (40\% by weight in water) at a temperature of $80^{\circ}$C, then removes the exposed silicon at a rate of $60\,\mu$m/hour in the direction perpendicular to the surface. The etch rate is anisotropic, exposing the \{111\} planes to create square pyramidal pits in the surface of the silicon wafer, as shown in Fig.\,\ref{fig:image1_pic}. The length of one side of the base is $\sqrt{2}$ times the depth of the pyramid. A temperature-stabilized water bath and re-condenser are used because the etch rate is extremely sensitive to small variations in the concentration and temperature of the etchant.  We have produced a number of pyramids in this way, ranging in depth from $0.9\,$mm to $3\,$mm. At the longest etching times, of $\sim 50$ hours, we found that the chips were damaged from the back where the etchant had managed to penetrate the silicon nitride, probably through scratches produced during handling and mounting. This problem was solved by covering the back side with a PTFE disk during etching.

Figure\,\ref{fig:image1_pic} is a photograph of a $2.8\,$mm-deep pyramid. We have added white lines to the image (here and in Figure\,\ref{fig:image2_pic}) in order to delineate the structure more clearly. The pyramid is perfectly formed except for the surface quality of the facets, which are locally smooth but exhibit roughness that is correlated over very long distances.  This can be seen on the illuminated facet in Fig.\,\ref{fig:image1_pic} and is even more obvious in the optical microscope image of Fig.\,\ref{fig:image1_mic}. This roughness is thought to be due to hydrogen bubbles, formed in the reaction between KOH and silicon, which modify the local silicon etch rate \cite{louro01} causing craters.  The structures we see are large because the etch time is long. A pyramid in this condition does not work as a MOT.

We tried a number of approaches to reduce the formation of this roughness during the etching. Adding a small amount of isopropanol as a surfactant produced no evident improvement in the surface quality. It has been suggested that bubbling air through the etchant can improve the surface because the increased oxygen concentration may help to remove the hydrogen \cite{louro01}. We tried this and found that it does indeed improve the surface quality. Figure\,\ref{fig:image2_pic} shows a pyramid that was made in this way, and the microscope image in Figure\,\ref{fig:image2_mic} shows the improvement in surface quality. The same improvement is found when we agitate the etchant with a magnetic stirrer or shake the sample to remove the hydrogen bubbles as they form. It is possible therefore that the effect of bubbling the air is primarily mechanical rather than chemical. Despite the improvement, this smoother pyramid is still not able to operate as a MOT.

Faced with this persistent long-range roughness, we considered how the surface might be smoothed by a second stage of processing. Small-scale roughness can be covered up by coating the surface with amorphous silicon dioxide, which becomes smooth after annealing \cite{lee01}. However, this approach does not solve our problem because the thickness of the covering would need to be many tens of $\mu$m in order to become comparable with the correlation length of the roughness.
A better approach is to etch the rough surfaces isotropically. We explored two isotropic etching techniques, (i) a wet etch using a mixture of hydrofluoric and nitric acids dissolved in acetic acid (HNA) and (ii) an inductively coupled plasma (ICP) etch using sulphur hexaflouride (SF$_6$).

The nitric acid in HNA reacts with silicon to produce SiO$_{2}$, which is then removed by reaction with the hydrofluoric acid. This polishing process produces a highly reflecting surface with no trace of roughness, but unfortunately it leaves the surfaces very rounded so that we no longer have the required pyramid geometry. Two different HNA solutions were tried, in the ratios vol\% (30:43:27) and (8:66:26), with etch rates differing by almost an order of magnitude.  These produced the same results, suggesting that the rounding is characteristic of the method and therefore that HNA polishing may not be suitable for our purposes.

We had more success with ICP etching, using a low-power plasma in pure SF$_{6}$ to produce fluorine radicals that attack the silicon surface.  We followed ``recipe ID 2" of Ref.\,\cite{larsen05}, which Larsen \textit{et al.} used to reduce the roughness of silicon micro-lens moulds. We measure the rate of the ICP etch to be $3\,\mu$m/min perpendicular to the surface. After 30\,min of etching, the surfaces are very much smoother and pyramids still retain their shape. Despite the large amount of material removed the faces remain flat and the edges are still sharp, as one can see in Figs. \ref{fig:image3_pic} and \ref{fig:image3_mic}. The quality of the surface continues to improve with ICP etching time, but we did not study this beyond $60\,$min.

In order to test whether a pyramid can operate as a MOT for ultra-cold atoms, we sputter-coat it with $25\,$nm of aluminum. This has a reflectivity of 0.8 at the operating wavelength, which allows the MOT to work without masking any part of the surface \cite{lewis09}. We were unable to obtain MOT operation with the unpolished pyramids, with or without agitation during the etching process. We believe this is due to the large scale roughness of the surface, which creates a coarse texture of the light intensity in the trapping region, having feature sizes comparable with the size of the atom cloud.  Pyramids that were polished in HNA do not work, even though the mirror surfaces are smooth, because the facets are curved. However, pyramids polished by ICP etching work well because the facets are flat with small scale surface roughness.

\begin{figure}[t]
\includegraphics[height=4.0cm]{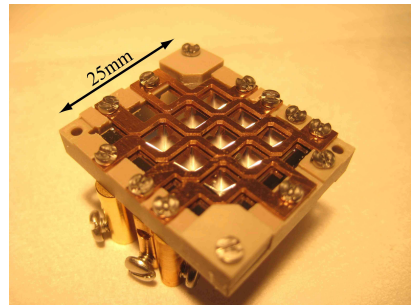}
\caption{\label{fig:RAY}The complete chip package consisting of a $3 \times 3$ array of silicon pyramid MOTs mounted in a rectangular PEEK holder $25 \times 30$ \,mm$^2$. The necessary magnetic fields are generated using a copper zig-zag array, which forms part of the package, above and below the chip.}
\end{figure}

The coated chips are mounted in a chip carrier machined from PEEK, a vacuum compatible polymer-glass mixture, as shown in Fig.\,\ref{fig:RAY}. Copper grids above and below the chip form current-carrying wires that generate the magnetic quadrupole fields required at the centre of each MOT. These are constructed by cutting patterns into a $0.7\,$mm-thick sheet using wire erosion. The vertical field gradient we use is typically $0.3\,$T/m, produced by currents of $\sim 3\,$A. Magnetic field coils can be integrated fully into the chip, as demonstrated in a previous paper \cite{lewis09}, but the present design is much more convenient for the purpose of testing pyramids. The entire assembly is housed in a high vacuum chamber where a small pressure (below $10^{-8}\,$mbar) of Rb atoms is provided by a current-activated dispenser. Additional bias coils external to the chip package are used to null the stray laboratory magnetic fields.

The experiment uses three diode lasers at $780\,$nm. A frequency-reference laser is locked by polarization spectroscopy to the $5S_\frac{1}{2}(F=2)\rightarrow5P_\frac{3}{2}(F'=3)$ transition of  $^{87}$Rb \cite{pearman02}. The MOT laser is frequency-locked $12\,$MHz below the $^{85}$Rb transition $5S_\frac{1}{2}(F=3)\rightarrow 5P_\frac{3}{2}(F'=4)$, 1.1\,GHz to the blue of the reference laser, using the ``side of filter" technique \cite{ritt04}. Atoms that fall into the $F=2$ ground state are re-pumped by a DAVLL-stabilized laser tuned to the $5S_\frac{1}{2}(F=2)\rightarrow5P_\frac{3}{2}(F'=3)$ $^{85}$Rb transition \cite{corwin98}. Each pyramid is  illuminated with $\sim 1\,$mW of circularly polarized MOT light and $\sim 50\,\mu$W of re-pump light.

\begin{figure}[t]
\includegraphics[height=4cm]{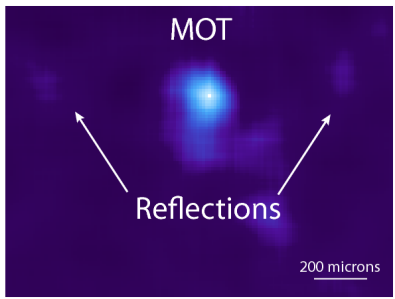}
\caption{\label{fig:MOT}Fluorescence image of $2000$ cold $^{85}$Rb atoms in a $2.5\,$mm-deep pyramid. Reflections of the MOT are just visible in two of the internal faces.}
\end{figure}

When cold atoms are trapped in a pyramid, they scatter enough MOT light to be detected by a CCD camera. Light is also scattered from the surfaces of the pyramid, but this background can be subtracted by taking images with and without atoms. Figure\,\ref{fig:MOT} shows such a difference image obtained with a $2.5\,$mm-deep pyramid.  In addition to the direct view of the cloud, one can see weaker images reflected from the sides of the pyramid. The light power detected by the camera can be converted to a number of atoms because we know the scattering rate per atom. In this case we measure $2000\pm400$ atoms. The density distribution is roughly Gaussian with a standard deviation of $80\,\mu$m in the two visible directions, parallel to the surface of the chip. The vertical size is $1/\sqrt{2}$ of this, giving a peak density of $4\times10^{14}$ atoms/m$^3$. A larger, $2.8\,$mm-deep pyramid yields $7000\pm 500$ atoms in a cloud of the same size, while a smaller $2.1\,$mm-deep pyramid has a cloud of $1000\pm300$ atoms and a similar radius. These MOTs all have a temperature of approximately $100\,\mu$K. In the rougher pyramids, where we could not detect a MOT, it is possible that there were some atoms, but the number would have had to be less then $300$ in order to escape detection

It is expected that the number of atoms should scale as $N\sim L^2 v_c^4$ \cite{lindquist92}, where $L$ is the size of the pyramid and $v_c$ is the maximum velocity of atoms that can be captured. Because these MOTs are small, the capture velocity is limited by the capture distance, which scales as $L$. The friction force $-\alpha v$ has a coefficient $\alpha$ that is almost constant over the capture region. Consequently $v_c\sim L$ and therefore $N\sim L^6$. Our observations are consistent with this scaling and inconsistent with the theory for larger MOTs, where the stopping distance is not the limiting factor and $N\sim L^{3.6}$ \cite{lindquist92}. With $L^6$ scaling, we expect a pyramid $1\,$mm deep to capture approximately $10$ atoms, which is below the detection limit of our current measurement scheme. We aim to work with such clouds by integrating optical micro-cavities into the same chip \cite{Trupke07} and to use them as a deterministic source of single atoms and photons.  There is also currently considerable interest in producing and trapping ultracold molecules on a chip. The integrated pyramid MOT offers an ideal starting point for this project since atoms are readily photo-associated in a MOT \cite{jones06}.

To summarize, we have shown how a pyramid MOT can be integrated into a silicon chip by a process of anisotropic wet etching, followed by polishing in an ICP etch. We have demonstrated that the integrated MOT works, provided the polishing is good enough, and we have compared a range of sizes to test the scaling law. This new atom trapping method operates with just one input beam of light and provides a simple way to integrate several atom sources on the same chip. It represents a substantial advance in atom chip technology and opens the way to new functionality for atom chips including integrated single atom or photon sources and molecules on a chip.

We are indebted to M. Trupke for discussions, to M. Tarbutt for suggesting the photoassociation of molecules in a pyramid MOT, and to S. Etienne \& T. Matsuura at the London Centre for Nanotechnology for fabrication assistance. This work is supported by the E. C's Seventh Framework project 216744 (CHIMONO), the UK EPSRC and QIPIRC, and the Royal Society.
%\bibliographystyle{apsrev}
%\bibliography{Bibliography}

\end{document}